\documentclass[journal=jacsat,manuscript=article]{achemso}
\usepackage{chemformula} 
\usepackage[T1]{fontenc} 

\usepackage{graphicx}
\usepackage{color}
\usepackage{dcolumn}
\usepackage{bm}
\usepackage{hyperref}
\usepackage[nolist,nohyperlinks]{acronym}
\usepackage{CJK}
\usepackage{braket}
\usepackage{nicefrac}
\usepackage{amsmath}
\definecolor{cblue}{RGB}{19,107,192}
\hypersetup{colorlinks=true,linkcolor=cblue,citecolor=cblue,urlcolor=cblue}
\usepackage{soul}

\newcommand{\kagome}{kagome}
\newcommand{\FeCoSn}{(Co$_{1-x}$Fe$_{x}$)Sn}
\newcommand{\FeCoSnN}{(Co$_{0.9}$Fe$_{0.1}$)Sn}
\newcommand{\FeCoSnF}{(Co$_{0.45}$Fe$_{0.55}$)Sn}
\newcommand{\FeCoSnCR}{(Co$_{0.6}$Fe$_{0.4}$)Sn}
\newcommand{\hex}{$P6/mmm$}
\newcommand{\magsg}{$P_{c}6/mcc$ (BNS 192.252)}
\newcommand{\orthosg}{$Cmmm$}

\newcommand{\GMmode}{$\Gamma_{5}^{+}$}

\title{Simultaneous development of antiferromagnetism and local symmetry breaking in a \kagome{} magnet \FeCoSnF{}}

\author{Tsung-Han Yang}
\affiliation{Neutron Scattering Division, Oak Ridge National Laboratory, Oak Ridge, Tennessee 37831, USA}

\author{Shang Gao}
\affiliation{Material Science \& Technology Division, Oak Ridge National Laboratory, Oak Ridge, Tennessee 37831, USA}

\author{Yuanpeng Zhang}
\affiliation{Neutron Scattering Division, Oak Ridge National Laboratory, Oak Ridge, Tennessee 37831, USA}

\author{Daniel Olds}
\affiliation{Photon Sciences Division, Brookhaven National Laboratory, Upton, New York 11973, USA}

\author{William R. Meier}
\affiliation{Material Science \& Technology Division, Oak Ridge National Laboratory, Oak Ridge, Tennessee 37831, USA}

\author{Matthew B. Stone}
\affiliation{Neutron Scattering Division, Oak Ridge National Laboratory, Oak Ridge, Tennessee 37831, USA}

\author{Brian C. Sales}
\affiliation{Material Science \& Technology Division, Oak Ridge National Laboratory, Oak Ridge, Tennessee 37831, USA}

\author{Andrew D. Christianson}
\email{christiansad@ornl.gov}
\affiliation{Material Science \& Technology Division, Oak Ridge National Laboratory, Oak Ridge, Tennessee 37831, USA}

\author{Qiang Zhang}
\email{zhangq6@ornl.gov}
\affiliation{Neutron Scattering Division, Oak Ridge National Laboratory, Oak Ridge, Tennessee 37831, USA}

\date{\today}

\begin{document}

\begin{tocentry}

 \begin{center}
     \includegraphics{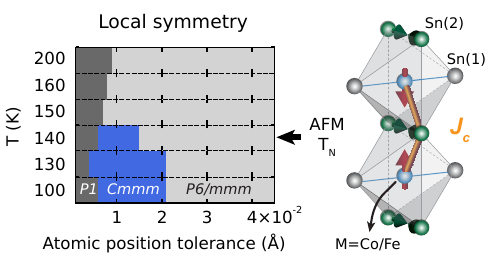}
 \end{center}
 
%
 

\end{tocentry}
\begin{abstract}

CoSn and FeSn, two \kagome{}-lattice metals, have recently attracted significant attention as hosts of electronic flat bands and emergent physical properties. However, current understandings of their physical properties are limited to the knowledge of the average crystal structure. Here, we report the Fe-doping induced co-emergence of the antiferromagentic (AFM) order and local symmetry breaking in \FeCoSnF{}. Rietveld analysis on the neutron and synchrotron x-ray diffraction data indicates A-type antiferromagnetic order with the moment pointing perpendicular to the kagome layers, associated with the anomaly in the $M$Sn(1)$_{2}$Sn(2)$_{4}$ ($M=$~Co/Fe) octahedral distortion and the lattice constant $c$. Reverse Monte Carlo (RMC) modeling of the synchrotron x-ray total scattering results captured the subtle local orthorhombic distortion involving off-axis displacements of Sn(2). Our results indicate that the stable hexagonal lattice above $T_{N}$ becomes unstable once the A-type AFM order is formed below $T_{N}$. We argue that the local symmetry breaking has a magnetic origin since the spatially varied $M$–Sn(2) bond lengths arise from out-of-plane magnetic exchange coupling $J_{c}$ via the exchange pathway $M$–Sn(2)–$M$. Our study provides comprehensive information on the crystal structure in both long-range scale and local scale, unveiling unique coupling between AFM order, octahedral distortion, and hidden local symmetry breaking.\footnote{Copyright  notice: This  manuscript  has  been  authored  by  UT-Battelle, LLC under Contract No. DE-AC05-00OR22725 with the U.S.  Department  of  Energy.   
The  United  States  Government  retains  and  the  publisher,  by  accepting  the  article  for  publication, 
acknowledges  that  the  United  States  Government  retains  a  non-exclusive, paid-up, irrevocable, world-wide license to publish or reproduce the published form of this manuscript, 
or allow others to do so, for United States Government purposes.  
The Department of Energy will provide public access to these results of federally sponsored  research  in  accordance  with  the  DOE  Public  Access  Plan 
(http://energy.gov/downloads/doe-public-access-plan)}       

\end{abstract}

\clearpage 
\maketitle
\section{Introduction}

Exploring the relationship between macroscopic physical properties and structural complexity stands as a pivotal focus in the study of modern functional materials. Understanding certain magnetic phenomena and other emerging properties becomes difficult when restricted within the framework of the average crystal structure. This challenge arises from the presence of nanoscale inheterogeneities, correlation disorder, or local symmetry breaking states in materials~\cite{Keen2015,Zhu2021,Xiao2020}. Thus, acquiring insight into the local deviations from the average structure becomes critical for studying these complex materials, such as batteries~\cite{Naoko2029,Bareno2010}, ferroelectrics~\cite{Chong2012}, thermoelectrics~\cite{Hu2018,Tyson2009,Li2018,Jiang2023}, Weyl semimetals~\cite{Zhang2022}, superconductors~\cite{Frandsen2017,Wang2020,Upreti2022}, and strongly correlated electronic oxides~\cite{Goodwin2006,Louca2010, Lu2017,Allieta2012}.

\begin{figure}[t!]
    \includegraphics[width=0.75\linewidth]{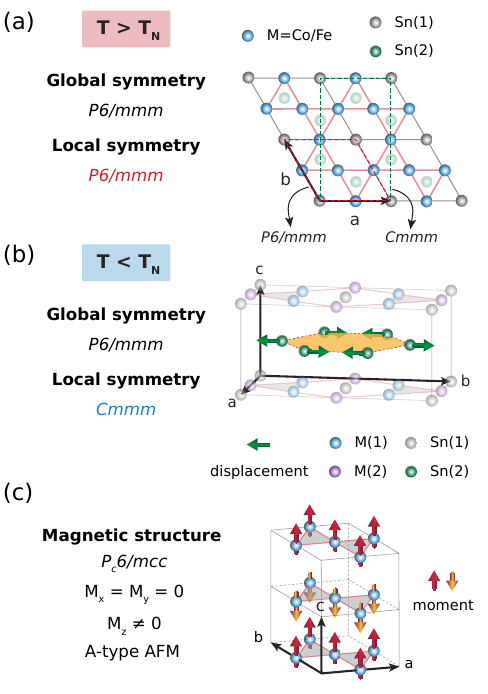}
\caption{Average crystal structure, local symmetry and magnetic structure of \FeCoSnF{}. (a) Top view of the kagome layer with both local and average symmetries being hexagonal \hex{} at $T>T_{N}$. $M$ and Sn(1) form the kagome layer, and Sn(2) atoms occupy the honeycomb layer between kagome layers. The red and green dashed lines show the relation between the hexagonal and orthorhombic cells. (b) The local symmetry is lowered to orthorhombic \orthosg{} in $T<T_{N}$ with unchanged average structure. The arrows represent the distortion of Sn(2) atomic positions in the honeycomb layer. (c) A-type AFM order in $T<T_{N}$ with magnetic space group \magsg{}.}
\label{fig:schematics}
\end{figure}

Kagome-lattice materials CoSn, FeSn and doped derivatives have attracted intense attention recently since they are ideal platforms to host novel magnetic or topological phases when electronic flat bands are coupled to spin-orbit couplings or magnetic order~\cite{Kang2020,Kang2020-2,Liu2020,Sales2021,Meier:2020}. The kagome lattice consists of corner-shared triangles, yielding intrinsic geometric frustration. CoSn and FeSn are isostructural and crystallize in a B35 hexagonal structure with space group \hex{} (No.~191) composed of Sn-filled Co/Fe-Kagome layer. As shown in Fig.~1(a), Co/Fe atoms form a Kagome lattice, and the Kagome layers are separated by Sn(2) honeycomb layers. CoSn is a Pauli paramagnet~\cite{Liu2020,Kang2020,Kang2020-2}, whereas FeSn exhibits A-type antiferromagnetic (AFM) order with planar moment below $T_{N}=$~365~K~\cite{YAMAGUCHI1967}. Previous theoretical and experimental studies show that the electronic flat bands of CoSn and FeSn are located hundreds meV away from the Fermi level~\cite{Kang2020,Kang2020-2,Multer:2023,Moore2022}. The substitution of Co by Fe atoms in \FeCoSn{} modifies the average number of electrons in the material, significantly tuning the Fermi level and magnetic properties~\cite{Meier2019}. When $x$ is increased to $\approx$~0.55, a new magnetic phase with out-of-plane moment was proposed by Meier \textit{et al.}~\cite{Meier2019} below $T_{N }=$~140~K, though the spin arrangements and moment size were not determined. To date, current understandings of CoSn, FeSn and their doped derivatives only relied on the framework of globally average structure~\cite{Kang2020,Kang2020-2,Liu2020,Sales2021,Meier2019,Multer:2023}. The local symmetry breaking remains elusive and its relationship to the magnetism and average structure remains mysterious in these materials.  

Here, we report the discovery of local symmetry breaking in \FeCoSnF{} that co-emerges with the AFM order below $T_{N}\approx$~140~K, by a combined use of synchrotron x-ray total scattering and neutron diffraction. Upon cooling, while the average B35 hexagonal structure remains unchanged, one anomaly was observed in lattice constant $c$ and $M$Sn(1)$_{2}$Sn(2)$_{4}$ ($M=$~Co/Fe) octahedral distortion at $T_{N}$. Local symmetry breaking occurs from hexagonal \hex{} to orthorhombic \orthosg{} below $T_{N}$, with striking off-axis Sn(2) displacements to a direction perpendicular to the crystalline $a/b$ direction, based on the reverse Monte Carlo (RMC) modeling on synchrotron x-ray total scattering results. The magnetic structure was determined to be A-type AFM order with moments of 0.59(8) $\mu_{B}$ per Co/Fe along the $c$–axis, where the magnetic moments are treated uniformly on the Fe/Co sites, based on the Rietveld analysis againsts the neutron diffraction patterns. The competition between the average hexagonal structure and local orthorhombic structure indicates the existence of lattice instability, with structural disorder forming when the AFM order is established. The hitherto overlooked local symmetry breaking has a magnetic origin, manifested primarily by the locally spatial variations in the (Co/Fe)–Sn(2) bond lengths and (Co/Fe)–Sn(2)–(Co/Fe) bond angles, which may be driven by magnetic exchange coupling via (Co/Fe)–Sn(2)–(Co/Fe) exchange pathway.

\section{Experimental and Computational Methods}

\subsection{Sample preparation and neutron powder diffraction and total scattering measurements}
\FeCoSnF{} and \FeCoSnN{} single crystals were synthesized using the flux growth method and the details have been reported in the previous work \cite{Meier2019}. The powder sample was prepared by pulverizing single crystals for measurements. We performed neutron powder diffraction measurements on the time-of-flight (ToF) diffractometer POWGEN at Spallation Neutron Source, Oak Ridge National Laboratory. \FeCoSnF{}, with a mass 4.8~g were loaded in a vanadium can and mounted on an orange cryostat. The neutron diffraction patterns at 300 K and 1.7 K were collected using the ToF neutron frame 2 with center wavelength of 1.5 ~\AA{}. To identify the possible structural or magnetic transitions in long-range scale, we conducted isothermal measurements of the neutron diffraction patterns using the highest-resolution neutron frame 3 with a center wavelength of~2.665~\AA{} at POWGEN, after 8-minute waiting time at each temperature to ensure thermal equilibrium. The \FeCoSnN{} sample with the mass of around 5 g was loaded in a POWGEN automatic changer (PAC) to cover temperature region of 10 to 300 K. We conducted neutron diffraction and pair distribution function (PDF) measurements using the neutron frame 1 with a center wavelength of~0.8~\AA{} on \FeCoSnN{}.  The data reduction for both diffraction and PDF patterns was performed using \textsf{MANTID} analysis package \cite{Arnold:2014}. The background signal was subtracted using the empty sample can measurement and the subtracted data was normalized against the vanadium rod signal to obtain diffraction data. A Q$_{max}$ of ~35~\AA$^{-1}$ was used to generate neutron PDF patterns. Rietveld refinements of the nuclear and magnetic structures for neutron diffraction data were performed using \textsf{GSAS-II} package \cite{GSAS2}. The symmetry-allowed magnetic structures were analyzed using the \textsf{Bilbao Crystallographic Server} \cite{BilbaoMaxMag}.

\subsection{Synchrotron x-ray total scattering measurements}
We performed synchrotron x-ray diffraction and pair distribution function (PDF) measurements on 28-ID-1 (PDF) beamline at NSLS-II, Brookhaven National Laboratory. A 30 mg polycrystalline sample from crushed single crystals was loaded in an I.D.$=$ 1~mm kapton capillary and sealed with epoxy. The temperature of the sample was controlled by a nitrogen cryostream, and data was collected between 100~K and 500~K. Measurements were carried out using an x-ray energy of 74.46~keV (0.1665~\AA{}). The amorphous silicon PerkinElmer area detector was used with the sample to detector distance of 1008~mm for conventional diffraction and 204~mm for PDF. A Ni standard was used to precisely determine the sample position for data reduction. The background was subtracted using a scan of an empty kapton capillary. The diffraction frame images were reduced to one dimension using \textsf{PyFAI} package \cite{Ashiotis:2015} and converted to PDF $G(r)$ (in \AA{}$^{-2}$) using \textsf{PDFgetX3} \cite{pdfgetx3} package with Q$_{max}=$~26~\AA$^{-1}$. The symmetry-adopted small box analysis was performed using \textsf{PDFgui} package~\cite{Farrow:2007}.

\subsection{Reverse Monte Carlo modeling}
We used \textsf{RMCprofile} software suite \cite{Tucker:2007,Zhang:2020} to perform the reverse Monte Carlo (RMC) modelings against x-ray $F(Q)$ and $G_{K}(r)$ simultaneously to analyze local distortions. A configuration of 7,200 atoms in a 10~$\times$~10~$\times$~12 supercell was used, with Fe dopants randomly distributed on Co sites. The $F(Q)$ was fitted in the range of 0.5 to 26.0 \AA{}$^{-1}$ in reciprocal space, and the $G_{K}(r)$ was simultaneously fitted from 2.42 to 25.0 \AA{} in real space. An instrumental broadening of 0.0363~\AA{}, obtained from the small-box analysis, was used in the modelings to avoid fitting nonphysical features in experimental data. It is worth mentioning that the definition of $G_{K}(r)$ (in barn) differs slightly from $G(r)$ (in \AA{}$^{-2}$) used in the small-box community. A detail comparison is provided by D. A. Keen~\cite{Keen:2001}. The $R$-value, $R \equiv \sqrt{\sum_{i}{|G_{fit}(r_{i})-G_{obs}(r_{i})|^{2}}/\sum_{i}{|G_{obs}(r_{i})|^{2}}}$, for each fit was minimized using the Metropolis algorithm. After achieving the converged fitting quality of $R~\approx{}$~0.01, we collapsed the supercell into a single unit cell and calculated the local Sn(2)–Sn(2) bond length in the folded structure for each temperature. The \textsf{FINDSYM} package \cite{Stokes:2005} was used to search for allowed symmetry operations of collapsed unit cell based on the atomic displacement tolerances for the local structure. The structural distortion modes contributing to local symmetry breaking were analyzed using the \textsf{AMPLIMODES} package on \textsf{Bilbao Crystallographic Server} \cite{BilbaoMaxMag,Perez:2010}.

\section{Results}
 
\subsection{Average crystal structure and magnetic order}
\begin{figure*}[t!]
    \includegraphics[width=1\linewidth]{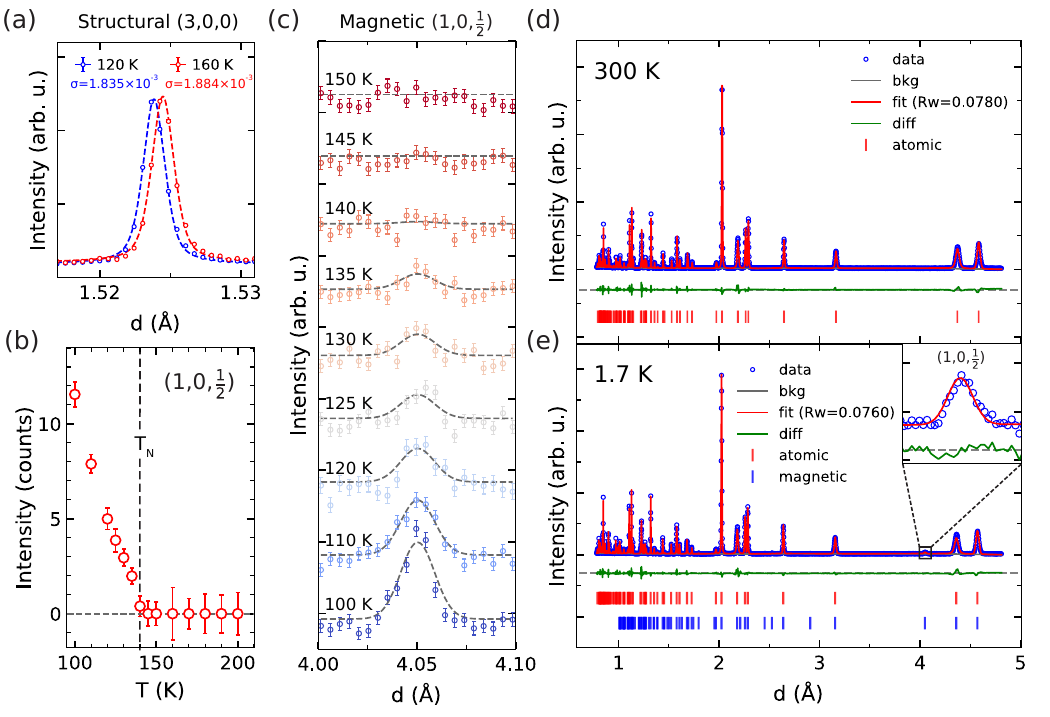}
\caption{Neutron diffraction results on a polycrystalline \FeCoSnF{} sample. (a) The structural (3,0,0) peak measured above and below T$_{N}$. The peak widths ($\sigma$) were extracted using a fit to a pseudo-Voigt function. No observable peak broadening or shape changes are found across T$_{N}$, suggesting the absence of a global structural phase transition. (b) Temperature dependence of magnetic (1,0,$\frac{1}{2}$) peak intensity. (c) Temperature dependence of neutron diffraction measurements for the magnetic (1,0,$\frac{1}{2}$) peak. The gray dashed lines represent fitted curves used to extract peak intensities. (d) and (e) show Rietveld refinements against powder neutron diffraction patterns of \FeCoSnF{} at 300~K and 1.7~K, respectively. The inset shows the zoomed experimental and fitted magnetic (1,0,$\frac{1}{2}$) peak.}
\label{fig:Mag_phase}
\end{figure*}

We begin by investigating the average crystal and magnetic structure of polycrystalline \FeCoSnF{} using temperature-dependent neutron diffraction. The sample quality and doping concentration are confirmed by the Rietveld refinement against the neutron diffraction data at 300~K, as shown in Fig.~\ref{fig:Mag_phase}~(d). The neutron scattering cross-sections of Fe and Co are 9.45 and 2.49 barn, respectively, providing sufficient contrast to distinguish the occupancies of these two magnetic ions. The high fidelity of the refinement ($R=$~0.0780) confirms the single \FeCoSn{} phase and yields an the Fe concentration of $x=$~0.558(45), closely matching the nominal value of $x_{nom}=$~0.55. The crystal structure of Fe-doped \FeCoSnF{} remains the hexagonal \hex{} symmetry, similar to the parent compounds CoSn or FeSn. The Fe dopants occupy only the Co-sites and are randomly distributed, as evidenced by our refinements. We obtain the lattice parameter $a=$~5.291(3)~\AA{} and $c=$~4.371(5)~\AA{}, with $c/a=$~0.826, larger than the value of parent compound CoSn ($c/a\approx$0.807)~\cite{LARSSON199679} as expected due to the larger atomic radius of Fe atom.

For $T<T_{N}$, our neutron diffraction measurements show that the average crystal structure of \FeCoSnF{} retains hexagonal \hex{} symmetry. The highest instrumental resolution of POWGEN ($\Delta d/d$) is approximately 0.0008. Even at this high resolution, no signs of peak splitting or broadening indicative of global structural phase transition were observed in POWGEN neutron diffraction patterns below $T_{N}$. For example, Fig.~\ref{fig:Mag_phase}~(a) presents a direct comparison of nuclear (3,0,0) peak at 120 K (below $T_{N}$) and 160 K (above $T_{N}$). The full width at half maximum (FWHM) of the peak are comparable at both temperatures. The synchrotron x-ray diffraction did not detect any peak splitting or broadening either in  $T<T_{N}$. In addition, small-box fittings to the PDF patterns in  $T<T_{N}$ indicates that the hexagonal symmetry provides a good fit to the data in the high-$r$ region (see Fig.~S1 in the Supporting Information (SI)). This indicates that the average crystal symmetry remains hexagonal in the long-range scale. All these results indicate that there is no global structural transition in long-range scale.

However, new magnetic Bragg peaks such as (1,0,$\dfrac{1}{2}$), (1,1,$\dfrac{1}{2}$) and (2,0,$\dfrac{1}{2}$), among others, appeared upon cooling, confirming the occurrence of single magnetic phase transition. The high resolution of (1,0,$\dfrac{1}{2}$) magnetic peak at different temperatures is shown in Fig.~\ref{fig:Mag_phase}~(c). The magnetic peak intensity of (1,0,$\dfrac{1}{2}$) starts developing at 140~K, and gradually increases with cooling as shown in Fig.~\ref{fig:Mag_phase}~(b). This confirms $T_{N}$=140 K, consistent with the value determined by the susceptibility measurements in previous report\cite{Meier2019}. All magnetic peaks can be indexed by a magnetic propagation vector $\vec{\textbf{\textit{k}}}=(0,0,\frac{1}{2})$. The magnetic moments are treated uniformly on the Fe/Co sites, which occupy the $3f$ Wyckoff positions (WP) in the primitive cell of \hex{} structure. 

Symmetry-allowed magnetic space groups are determined using the \textsf{Bilbao Crystallography Server} \cite{BilbaoMaxMag} on \FeCoSnF{}. All the planar magnetic orders and the G-type axial AFM order (moments order antiferromagnetically in both in-plane and out-of-plane) are ruled out by the inconsistency of relative magnetic Bragg's peak intensities between the model and the experimental data. The determined magnetic structure is A-type axial antiferromagnetic order with the magnetic space group \magsg{}, in which the magnetic moments are parallel within the \kagome{} plane and antiparallel between \kagome{} layers [Fig.~\ref{fig:Mag_phase}~(e)]. The refined magnetic moment is 0.59(8) $\mu_{B}$ on each Fe/Co site.

\subsection{Lattice anomaly and octahedral distortion at $T_{N}$}
\begin{figure} 
    \includegraphics[width=0.6\linewidth]{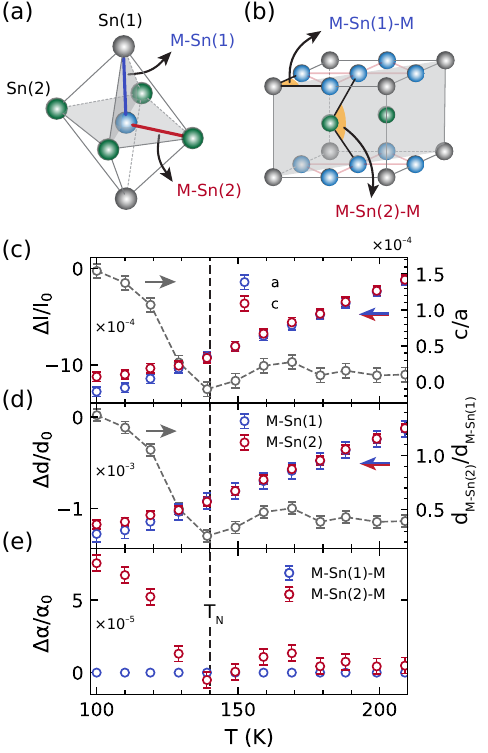}
\caption{Temperature dependence of the structural parameters in long-range scale extracted from Rietveld refinements on powder x-ray diffraction patterns against \hex{} hexagonal structure in \FeCoSnF{}. (a) Schematic for octahedral coordination and definitions for $M$-Sn(1) and $M$-Sn(2) bonds. (b) The definition of $M$–Sn(1)–$M$ and $M$–Sn(2)–$M$ bond angles. Temperature dependence of (c) lattice parameters $a$, $c$ and $c/a$, (d) long-range bond lengths for $M$–Sn(1), $M$–Sn(2) ($M=$~Co/Fe) and octahedral distortion parameter d($M$–Sn(2))/d($M$–Sn(1)), and (e) bond angles of $M$–Sn(1)–$M$ (in-plane) and $M$–Sn(2)–$M$ (out-of-plane). All the refined parameters are scaled by the values at 220~K for better comparison. }
\label{fig:TdepLatt}
\end{figure}

To explore the possible correlation between spins and lattice, we performed Rietveld refinements against synchrotron x-ray diffraction (XRD) data to examine the temperature-dependent crystal structure on the long-range scale. While the in-plane lattice constant $a$ shows a monotonic decrease upon cooling, an anomaly with a slope change is observed at $T_{N}$ in the out-of-plane lattice constant $c$, as shown in Fig.~\ref{fig:TdepLatt}~(c). Correspondingly, there is an anomalous increase in the $c/a$ below $T_{N}$. 

In \FeCoSnF{}, each Co/Fe atom is coordinated by two Sn(1) atoms and four Sn(2) atoms, forming an octahedral geometry. At 300 K, the Sn(1)–$M$–Sn(2) ($M=$~Fe/Co) angles retain at 90$^{\circ}$, while the Sn(2)–$M$–Sn(2) ($M=$~Fe/Co) angles significantly deviate to 69.88$^{\circ}$ and 110.12$^{\circ}$. The octahedral distortion parameter can be characterized by the ratio of the bond lengths $M$–Sn(2) to $M$–Sn(1) ($M=$~Co/Fe)~\cite{Katsufuji2008}. The temperature dependence of the bond lengths $M$–Sn(2), $M$–Sn(1) and the octahedral distortion parameter $d_{M-Sn(2)}/d_{M-Sn(1)}$ based on our Rietveld analysis on XRD patterns is shown in Fig.~\ref{fig:TdepLatt}~(d). There is no anomaly in the $M$–Sn(1) bond length, but we observe a slope change in the $M$–Sn(2) bond length and a rapid increase of the octahedral distortion parameter, specifically the suppression of octahedron along the $M$-Sn(1) direction, below $T_{N}$. The temperature dependence of the in-plane $M$–Sn(1)–$M$ and out-of-plane $M$–Sn(2)–$M$ bond angles are summarized in Fig.~\ref{fig:TdepLatt}~(e). There are clearly enlarged $M$–Sn(2)–$M$ angles along the $c$ direction below $T_{N}$, whereas no anomaly is observed in the $M$–Sn(1)–$M$ angles. 

The anomalies in the lattice constant $c$, bond length $M$–Sn(2) and bond angle $M$–Sn(2)–$M$ in the long-range scale, as well as the octahedral distortion parameter at $T_{N}$ provide compelling evidence of a strong interplay between the spin and lattice degrees of freedom in \FeCoSnF{}.  


\subsection{Local symmetry breaking at $T_{N}$}
\begin{figure*}[t!]
    \includegraphics[width=1\linewidth]{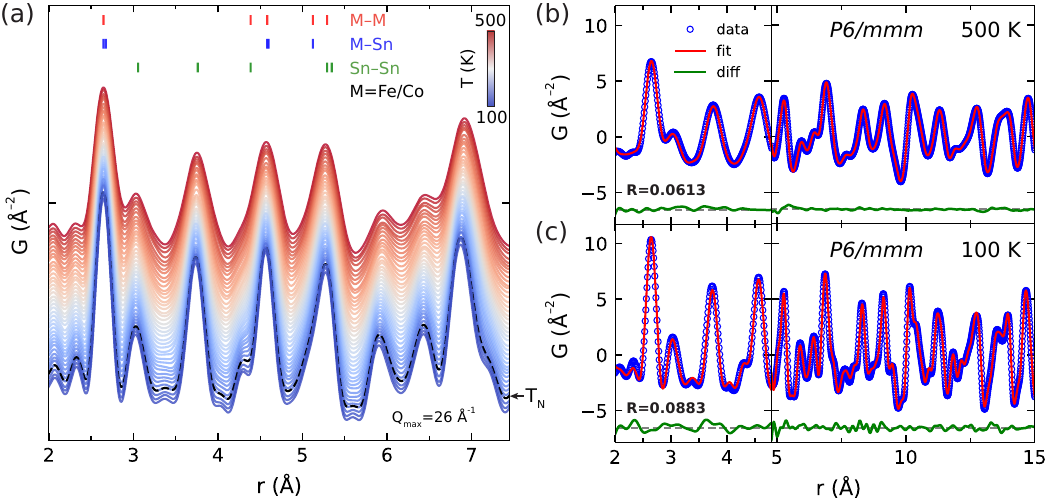}
\caption{Temperature-dependent synchrotron x-ray PDF patterns and small-box refinements against \hex{} symmetry in \FeCoSnF{}. (a) Temperature dependence of the synchrotron x-ray PDF patterns. The interatomic pair distances labeled by color ticks are calculated from the hexagonal \hex{} structure. The dashed curve indicates the PDF at $T_{N}$. (b) and (c) are fitted PDFs using \textsf{PDFgui} package at 500~K and 100~K, respectively. The vertical line indicates the boundary of the first unit cell. }
\label{fig:xPDFWaterfall}
\end{figure*}

Motivated by the observation of notable spin-lattice coupling in \FeCoSnF{}, we now go beyond the long-range scale to investigate the evolution of the local structure across the magnetic transition. Figure~\ref{fig:xPDFWaterfall}~(a) shows the temperature-dependent synchrotron x-ray pair distribution function (PDF) patterns, where the color ticks indicate the corresponding interatomic pair distances calculated using the high symmetry hexagonal \hex{} structure. We first perform a small box analysis to examine the local structure in detail using \textsf{PDFgui} \cite{Farrow:2007}. The high-quality refinement at 500~K indicates a balanced concentration of Fe and Co does not introduce the local distortion in \FeCoSn{}, while the \kagome{} structure remains undistorted [Fig.~\ref{fig:xPDFWaterfall}~(b)]. In contrast, the refinement for the 100~K cannot achieve the comparable fitting quality using the same hexagonal \hex{} structure, with the deviation even more pronounced within the $r$-range of the first unit cell. The degradation in the fit quality starts at $T_{N}$ [Fig.~S2 in the SI], indicating that hexagonal symmetry is insufficient to accurately describe the local structure.

\begin{figure*}[t!]
    \includegraphics[width=1\linewidth]{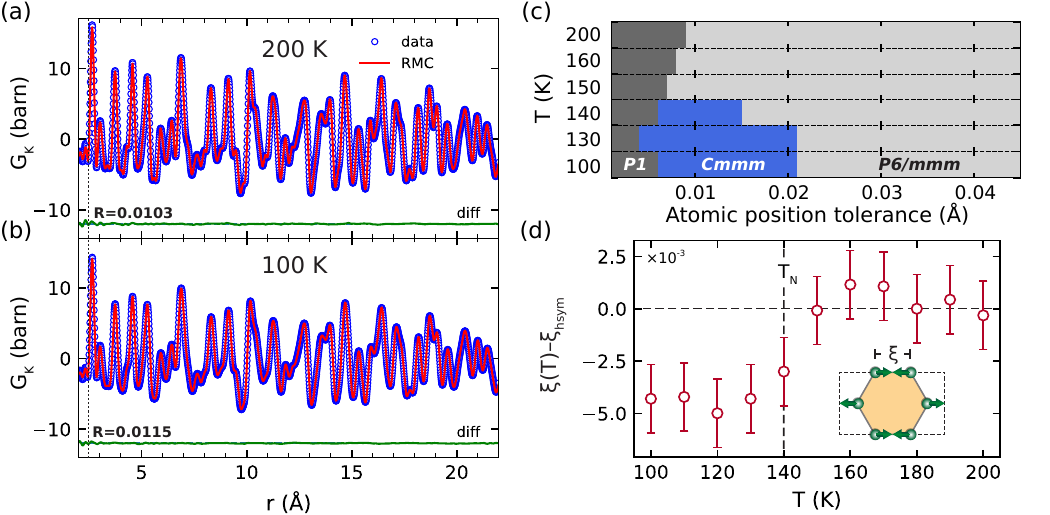}
\caption{Fitted PDF patterns $G_{K}(r)$, local symmetry and distortion extracted from the large-box analysis using RMC modeling on \FeCoSnF{}. Fitted results for $G_{K}(r)$ at (a) 200~K and (b) 100~K. The vertical dashed line indicates the \textit{r} minimum for the fit. (c) Temperature dependence of the space group determination using \textsf{FINDSYM} package. The orthorhombic symmetry in local scale starts developing at $T_{N}$ and persists down to lower temperatures. The symmetry returns to $P1$ at small atomic position tolerance. (d) Temperature dependence of the local Sn(2)–Sn(2) distance $\xi$ (in fractional units) within the collapsed unit cell. $\xi_{hsym}$ represents the fractional Sn(2)–Sn(2) distance in the high-symmetry hexagonal structure. Error bars are estimated from error propagation, based on the statistical deviation of atomic positions in the supercell.}
\label{fig:RMC}
\end{figure*}

To identify the possible local distortion, we performed the reverse Monte Carlo (RMC) modeling to track the temperature evolution of structural symmetry across the magnetic transition. A 10$\times$10$\times$12 supercell is used to enable the real-space fit range up to 25 Å and the \textsf{RMCprofile} simultaneously fits the synchrotron x-ray total scattering data in real ($G_{K}(r)$) and reciprocal ($F(Q)$) spaces using equal weights. Figure~\ref{fig:RMC}~(a) and (b) show the resulting real-space fits for $T>T_{N}$ and $T<T_{N}$. We performed RMC modeling for temperatures in the range of 100~K to 200~K, and the fitted structures were used for symmetry determination and local structure analysis.

To determine the local symmetry above and below $T_{N}$, we use \textsf{FINDSYM} package~\cite{Stokes:2005} to search for the symmetry of the local structure against the tolerances of atomic displacement. We collapse the 10$\times$10$\times$12 supercell to a 1$\times$1$\times$1 unit cell with $P1$ symmetry and then determine the space group and available cell based on allowed symmetry operations within the provided tolerances. Atomic position tolerance refers to the allowable deviation of atomic positions from their ideal symmetrical positions, the deviation less than the tolerance will be treated as zero. Therefore, a smaller tolerance tends to result in lower symmetry and vice versa. We sweep the atomic position tolerance value in the range of 0 to 0.045 \AA{} with 0.001 \AA{}/step for all analyses, generating a mapping of temperature-dependent local symmetry against atomic position tolerance, as presented in Fig.~\ref{fig:RMC}~(c).

At 100 K, the local orthorhombic distortion (\GMmode{} mode) with the symmetry of \orthosg{} (No.~65) is determined with the tolerance of atomic displacements up to 0.021 \AA{}. With increasing temperatures, the tolerance of atomic displacements corresponding to the local \orthosg{} symmetry decreases, indicating that the local orthorhombic distortion reduces at higher temperatures as expected. Such a temperature dependence of the tolerance for \orthosg{} symmetry provides strong evidence to verify the orthorhombic distortion on a local scale. Notably, the local symmetry \orthosg{} disappears and evolves to match the average symmetry \hex{} above $T_{N}$. Our results indicate a surprising co-emergence of the A-type AFM order and the local symmetry breaking at $T_{N}$.

\subsection{Local off-axis displacements of Sn(2) and lattice instability in $T<T_{N}$}
\begin{figure}[t!]
    \includegraphics[width=1\linewidth]{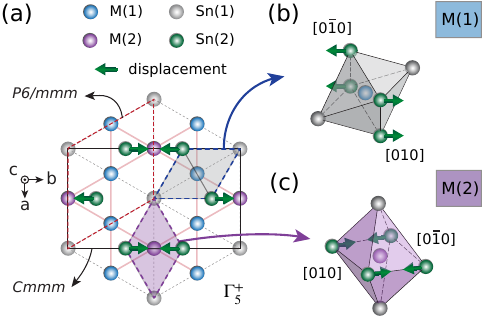}
\caption{Local distortion motif and local orthorhombic distortion in the orthorhombic notation of \FeCoSnF{}. The green arrows indicate the atomic displacements of Sn(2). (a) The single site on $3f$ WP in the hexagonal symmetry splits into $2b$ and $4e$ WPs in the orthorhombic symmetry, indicated by $M$(1) and $M$(2). The Sn(2) atoms in honeycomb layers move along orthorhombic $b$–axis. (b) and (c) show the distinct octahedral distortion for these two sites. The locations of two nonequivalent octahedra are shown in (a).}
 
\label{fig:PDFCmmm}
\end{figure}

The obtained Wyckoff positions and average atomic positions over the 10$\times$10$\times$12 supercell for the local orthorhombic structure \orthosg{} at 100 K are summarized in Table~\ref{tab:Cmmm}. The local orthorhombic distortion is characterized by an in-plane shift of Sn(2) atoms along the orthorhombic $b$ axis in the honeycomb layer between the kagome layers. Figure~\ref{fig:PDFCmmm}~(b) illustrates the orthorhombic local distortion in \FeCoSnF{} below $T_{N}$. The x position of Sn(2) does not shift along any crystalline axes of the hexagonal lattice, resulting in an off-axis displacements. Instead, it shifts to a direction perpendicular to one of the in-plane crystalline $a/b$–axis of the hexagonal unit cell, aligning along the new orthorhombic $b$–axis in the \orthosg{} local structure.  Due to these local atomic displacements of Sn(2), one Co/Fe octahedron in hexagonal symmetry splits into two nonequivalent Fe/Sn octahedral environments indicated by Fig.~\ref{fig:PDFCmmm}~(a)–(c). In local scale, distortion can occur along the $+b$ or $-b$ axis in varying magnitudes across different unit cells, indicating the presence of local site disorder. However, on a long-range scale, these orthorhombic distortions, differing in direction and magnitude, cancel each other out. Consequently, the average crystal structure can be represented by the hexagonal \hex{}. The correlation length of the local distortion is estimated to be approximately 25 \AA{}, as determined using a correlation function as shown in Fig.~S1 in the SI.

It is of interest to point out that the local atomic displacements of Sn(2) indicate the existence of locally varied $M$–Sn(2) bond lengths and $M$–Sn(2)–$M$ bond angles, and their average magnitudes in the long-range scale exhibit an anomaly at $T_{N}$ as shown in Fig.~\ref{fig:TdepLatt}~(d–e). Therefore, the local atomic displacements of Sn(2) are the microscopic origin of the anomalies of long-range $M$–Sn(2) bond lengths, $M$–Sn(2)–$M$ bond angles and the resultant lattice constant $c$. As shown in Fig.~\ref{fig:RMC}~(d), the temperature dependence of the local Sn(2)–Sn(2) bond lengths obtained from the RMC analysis in the collapsed supercell shows a clear increase below $T_{N}$, further confirming the co-emergence of A-type AFM order and the local orthorhombic distortion.

Above $T_{N}$, both the average and local symmetries are the same hexagonal structure with space group \hex{}, yielding a stable lattice. Below $T_{N}$, the mismatched average hexagonal symmetry and local orthorhombic symmetry indicate that the stable lattice becomes unstable below $T_{N}$ \cite{Allieta2012}. The lattice instability is primarily manifested by the presence of locally disordered Sn(2) atoms in a direction perpendicular to the hexagonal $a/b$–axis. 

\begin{table}[h]
\caption{Refined local structural parameters obtained from the RMC fit at 100~K. The local symmetry is orthorhombic \orthosg{} (No.~65) with  $a=$~5.269(4)~\AA{}, $b=$~9.127(6)~\AA{}, and $c=$~4.351(3)~\AA{}.}
\begin{tabular}{ c c c c c c}
Atom & WP & x/a & y/b & z/c & occ. \\
\hline 
Sn(1) & 2b & 0 & 0.5 & 0 & 1.0 \\
Sn(2) & 4h & 0 & 0.1654 & 0.5 & 1.0 \\
Fe(1) & 4e & 0.25 & 0.25 & 0 & 0.55 \\
Fe(2) & 2a & 0 & 0 & 0 & 0.55 \\
Co(1) & 4e & 0.25 & 0.25 & 0 & 0.45 \\
Co(2) & 2a & 0 & 0 & 0 & 0.45
\end{tabular}
\label{tab:Cmmm}
\end{table}


\section{Discussion}
\subsection{Fe-doping induced A-type AFM order and spin reorientation}

 CoSn is a Pauli paramagnet without a long-range magnetic order~\cite{Kang2020,Kang2020-2,Liu2020}. As the Fe concentration increases to $\approx$ 0.4–0.8, our neutron diffraction studies reveal that Fe doping induces an A-type antiferromagnetic (AFM) order with moments aligned along the $c$–axis below $T_{N}$. With further increasing Fe concentration to $\approx$ 0.8–0.97, an axial A-type AFM order was formed at high temperature, which transforms to a tilted AFM order at lower temperatures, showing the second magnetic transition \cite{Meier2019}. At higher Fe concentrations ($>$0.97), the axial AFM order disappears, leaving only the planar A-type AFM order below $T_{N}$, indicating a Fe-doping induced spin reorientation \cite{Meier2019}. Note that although the doping ratio alters magnetic structures and the lattice parameters, the average crystal symmetry remains hexagonal \hex{} for 0~$\leq x \leq$~1 \cite{Sales2021,Meier2019}. 

It is of interest to discuss why increasing Fe doping affects the magnetic order in \FeCoSn{}. In CoSn, relatively flat electronic bands are located $\sim$ 100 meV below the Fermi energy as seen from angle-resolved photoelectron spectroscopy (ARPES) and density functional theory (DFT) calculations \cite{Kang2020,Kang2020-2,Liu2020}. In \FeCoSn{}, iron has one less electron than Co, and doping Fe on the Co site can modify the electronic structure of the material. Such hole doping is expected to move the flat bands closer to E$_{F}$ and increase the density of states near the Fermi level, thus inducing an appearance of long-range AFM order in compounds when $x$ reaches around 0.4. The axial A-type AFM order in Co-rich \FeCoSnCR{} indicates that the preferred orientation of Co moments points to the $c$–axis. The small ordered moment of $\approx$ 0.59(8) $\mu_{B} $/(Co/Fe) in \FeCoSn{} indicates the electron itinerancy in these compounds. In the ending compound FeSn, the planar A-type AFM order shows that the preferred orientation of the Fe moments lies in the $ab$ plane, like many other compounds, for instance, Fe-based pnictides \cite{Cruz2008,Lester2009}. Thus, as Fe concentration further increases within $x\approx$ 0.8–0.97 in \FeCoSn{}, the competition of the distinct preferred orientations of the Co and Fe moments destabilizes the axial A-type AFM order present at high temperature region, leading to a tilted/canted AFM order at low temperatures as a more stable magnetic state. When $x$ exceeds $\approx$ 0.97, the dominant Fe ions result in a stable A-type AFM order with moments oriented in the $ab$ plane.

\subsection{Magnetic origin of the local symmetry breaking}
 
A common phenomenon in magnetic materials is that local symmetry breaking induced by chemical doping occurs at a higher temperature than the magnetic ordering temperature. For example, in Fe-based superconductor (Sr$_{0.71}$Na$_{0.29}$)Fe$_{2}$As$_{2}$ \cite{Frandsen2017}, local symmetry breaking from a tetragonal to an orthorhombic local structure happens at $\approx$ 300~K, much higher than the AFM transition temperature $\approx$ 140~K. In addition, the local distortions were found to appear with ferrimagnetic order in Fe$_{3}$O$_{4}$\cite{Perversi2019} or FM order in Co$_{3}$Sn$_{2}$S$_{2}$\cite{Zhang2022}, respectively, without chemical doping. In contrast, local symmetry breaking in Fe-doped \FeCoSnF{} co-emerges with the AFM order, which is associated with the octahedral distortion and anomaly in the out-of-plane lattice constant $c$ in the long-range scale. Note that we also investigated paramagnetic \FeCoSnN{}\cite{Meier2019} with low iron concentration using neutron diffraction and total scattering. We confirmed the absence of the magnetic order down to 10 K. Our PDF analysis indicates that the local symmetry is identical to the average hexagonal symmetry, with no evidence of local symmetry breaking [Fig.~S3 in the SI]. However, as the Fe concentration increases to 0.55 in \FeCoSnF{}, Fe doping induces the simultaneous development of the AFM  order and local symmetry breaking. All of these results suggest that Fe doping induces a new emergent coupling between magnetism and lattice at $T_{N}$ in \FeCoSnF{}.

\begin{figure}[t!]
        \includegraphics[width=0.9\linewidth]{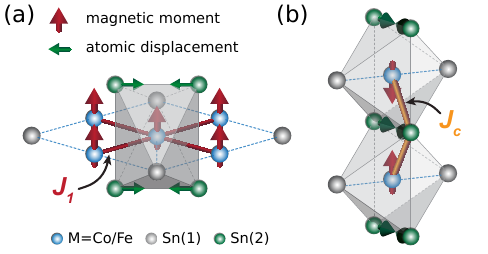}
\caption{Geometrical representation of (a) direct NN exchange interaction $J_{1}$ via Co/Fe–Co/Fe pathway and (b) superexchange interaction $J_{c}$ via Co/Fe–Sn(2)–Co/Fe pathway. The red arrows represent magnetic moments, and the green arrows indicate the local atomic displacements of Sn(2).}
\label{fig:J}
\end{figure}

Figure \ref{fig:J} illustrates the magnetic structure, chemical environments, and the important exchange couplings in \FeCoSnF{}. Along the $c$ direction, there are two nearest-neighbor (NN) $M$ ($M=$~Co/Fe) atoms surrounding one $M$ atom. Within the $ab$ plane, each $M$ atom is surrounded by 4 NN $M$ atoms with the same distance (see Fig.~\ref{fig:J}~(a)). The $M$Sn(1)$_{2}$Sn(2)$_{4}$ octahedra are edge shared along the $c$–axis but face-shared in the $ab$ plane. Given that \FeCoSnF{} and FeSn exhibit similar A-type AFM order, with the main difference being the preferred orientation of the moments, one would expect the magnetic exchange couplings to show similar characteristics. Inelastic neutron scattering on FeSn \cite{Xie:2021,Do:2022} has revealed two strongest magnetic exchange couplings: in-plane NN FM $J_{1}\approx-20.7$ meV and the out-of-plane AFM $J_{c}\approx 9.5$ meV. Thus, AFM $J_{c}$ is approximately 1/2 of the FM $J_{1}$, indicating a significant AFM interaction that stabilizes the antiparallel arrangement of spins to form the A-type AFM order. Despite the magnitudes of the $J_{1}$ and $J_{c}$ in \FeCoSnF{} being lower than those in FeSn, as manifested by the lower $T_{N}$ (140~K vs 365~K), these two magnetic exchange couplings should be the strongest ones with the same sign in \FeCoSnF{}.

Since the NN in-plane $M$–$M$ bond is the shortest one (2.638 \AA{}), the likely exchange coupling for $J_{1}$ is direct FM exchange interaction between face-shared octahedra via NN $M$–$M$ bond. The NN $M$–$M$ bond length scales with the lattice constant $a$ as depicted in Fig.~\ref{fig:TdepLatt}~(c), and therefore, there is no anomaly in the long-range $M$–$M$ bond length at $T_{N}$. In the local scale, our RMC analysis on synchrotron x-ray PDF patterns shows there are no displacements of $M$ atoms either below $T_{N}$. Therefore, $J_{1}$ may not be closely related to the local symmetry breaking.

As for $J_{c}$ along the $c$–axis, the direct $M$–$M$ distance is notably long ($\approx$~4.361~\AA{}), suggesting minimal direct exchange interaction. In contrast, the $M$–Sn(2) bond length within one octahedron tilting to the $c$ direction is much shorter at 2.659~\AA{}, comparable to the NN in-plane $M$–$M$ bond length ($\approx$~2.638~\AA{}). Thus, the AFM $J_{c}$ is expected to arise from super-exchange interaction between edge-shared octahedra via the $M$–Sn(2)–$M$ exchange pathway, with Sn(2) serving as the mediator. Furthermore, in the long-range scale, there exists octahedral suppression with enlarged $M$–Sn(2)–$M$ bond angles below $T_{N}$, as illustrated in Fig.~\ref{fig:TdepLatt}~(d–e). In the local scale, we observed local displacements of Sn(2) atoms indicative of the spatially varied M-Sn(2) bond lengths and M-Sn(2)-M angles below $T_{N}$, which breaks a local symmetry from hexagonal to orthorhombic. Therefore, the out-of-plane AFM interaction $J_{c}$ may be the driving force of the local symmetry breaking by inducing the locally disordered Sn(2) atoms via the $M$–Sn(2)–$M$ pathway.

\section{Summary and conclusion}

In summary, we report a Fe-doping driven co-emergence of local symmetry breaking, the A-type AFM order and anomalies in the octahedral distortion and lattice constant $c$ at $T_{N}$ in \FeCoSnF{}, highlighting a novel and strong coupling between spins and lattice across both long-range and local scales. The presence of locally disordered and off-axis Sn(2) atoms are the microscopic origin for anomalies in Co/Fe–Sn(2) bond length, Co/Fe–Sn(2)–Co/Fe bond angles, and lattice constant $c$ in the long-range scale. The stable lattice above $T_{N}$ becomes unstable upon cooling below $T_{N}$ due to the mismatched average hexagonal symmetry and local orthorhombic symmetry. The local symmetry breaking with disordered Sn(2) atoms is driven primarily by the out-of-plane magnetic interaction $J_{c}$ via the (Co/Fe)–Sn(2)–(Co/Fe) pathway. Theoretical calculations on \FeCoSnF{} are called to get more precise electronic structures, including flat bands, by considering the overlooked local orthorhombic structure. From an experimental perspective, this overlooked yet important local symmetry breaking with site disorder needs to be investigated in A-type planar AFM FeSn, its derivatives, and other related kagome magnets. 

\section{Acknowledgments} 
 This research used resources at the Spallation Neutron Source, a DOE Office of Science User Facility operated by the Oak Ridge National Laboratory. The beam time was allocated to POWGEN on proposal number IPTS-26683.1. Neutron scattering data collection was supported by the U.S. Department of Energy, Office of Science, Basic Energy Sciences, Materials Science and Engineering Division. This research used resources at the 28-ID-1 (PDF) beamline of the National Synchrotron Light Source II, a U.S. Department of Energy (DOE) Office of Science User Facility operated for the DOE Office of Science by Brookhaven National Laboratory under Contract DE-SC0012704.

 \section{Supporting Information Available} 

 Supplementary information: Supplementary figures on the estimated correlation length of the local distortion of \FeCoSnF{}, $R-$value of the Rietveld refinements on \FeCoSnF{}, and fitted neutron pair distribution patterns of \FeCoSnN{}.  

\bibliography{Refs_rev.bib}
\end{document}


\begin{figure*}[t!]
    \includegraphics{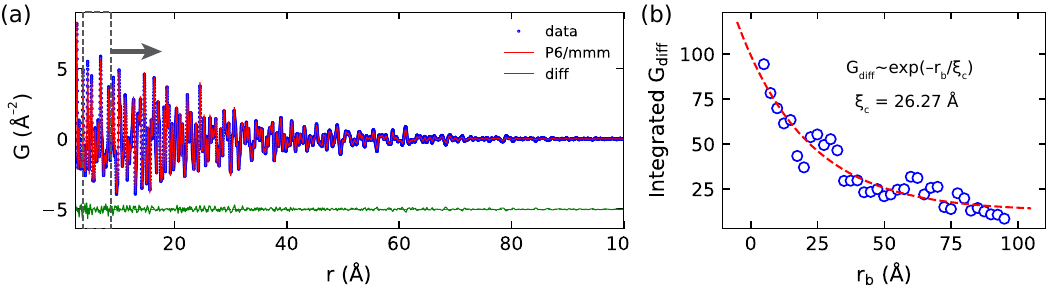}
\caption{ Estimated correlation length of the local distortion of \FeCoSnF{}. (a) Small-box fit at 100~K using the high-symmetry hexagonal \hex{} structure, with a fit range from 2.42 to 100~\AA{}. (b) The integrated $G_{diff}$ is calculated using a 5 \AA{} rolling-box as shown in (a), where $r_{b}$ represents the center of the rolling-box. To estimate the correlation length of the local distortion, the integrated $G_{diff}$ is fitted using the correlation function $\propto{}e^{-r_{b}/\xi_{b}}$. The estimated correlation length of the local structure is $\xi_{c}=$~26.27~\AA{}.}
\label{fig:S3}
\end{figure*}

\begin{figure*}[t!]
    \includegraphics{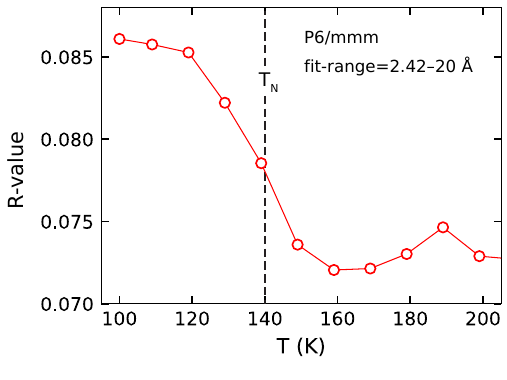}
\caption{ Temperature-dependent $R$-value from small-box analysis using hexagonal \hex{} structure of \FeCoSnF{}. The fitting r-range is from 2.42 to 20 \AA{}. The abrupt increase in the $R$-value at T$_{N} = $~140~K indicates the local symmetry is lower than the hexagonal \hex{} symmetry.}
\label{fig:S2}
\end{figure*}

\begin{figure*}[t!]
    \includegraphics{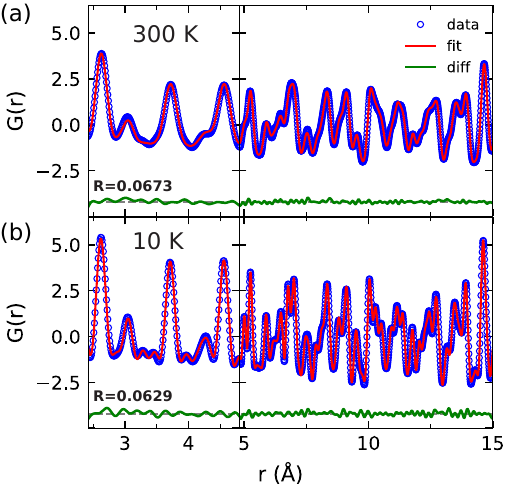}
\caption{Fitted neutron pair distribution patterns for nominal \FeCoSnN{} sample using PDFgui package at (a) 300~K and (b) 10~K. The small-box analysis was performed using hexagonal \hex{} structure, with a fit range from 2.42 to 15~\AA{}. The comparable fitting quality throughout entire fit range indicates no observable local distortion at 10~K and 300~K.}
\label{fig:S4}
\end{figure*}
